\documentclass[pra,showpacs,showkeys,amsmath,amssymb,superscriptaddress,preprintnumbers,floatfix, preprint]{revtex4}
\usepackage{bm,amsfonts, mathtools}
\usepackage{graphicx,color, wasysym}
\usepackage{textcomp}

\begin{document}
\renewcommand{\arraystretch}{2.3}

\title{The Ramanujan master theorem and its implications for special functions}

\author{K.~G\'{o}rska}
\email{kasia_gorska@o2.pl}

\affiliation{Laboratoire de Physique Th\'{e}orique de la Mati\`{e}re Condens\'{e}e (LPTMC), Universit\'{e} Pierre et Marie Curie, CNRS UMR 7600, Tour 13 - 5i\`{e}me \'{e}t., Bo\^{i}te Courrier 121, 4 place Jussieu, F 75252 Paris Cedex 05, France}

\affiliation{H. Niewodnicza\'{n}ski Institute of Nuclear Physics, Polish Academy of Sciences, ul. Eljasza-Radzikowskiego 152,  PL 31342 Krak\'{o}w, Poland}

\author{D.~Babusci}
\email{danilo.babusci@lnf.infn.it}

\affiliation{INFN - Laboratori Nazionali di Frascati, v. le E. Fermi, 40, IT 00044 Frascati (Roma), Italy}

\author{G.~Dattoli}
\email{dattoli@frascati.enea.it}

\affiliation{ENEA - Centro Ricerche Frascati, v. le E. Fermi, 45, IT 00044 Frascati (Roma), Italy}

\affiliation{Universit\'e Paris XIII, LIPN, Institut Galil\'{e}e, CNRS UMR 7030, 99 Av. J.-B. Clement, F 93430 Villetaneuse, France}

\author{G.~H.~E.~Duchamp}
\email{ghed@lipn-univ.paris13.fr}

\affiliation{Universit\'e Paris XIII, LIPN, Institut Galil\'{e}e, CNRS UMR 7030, 99 Av. J.-B. Clement, F 93430 Villetaneuse, France}

\author{K.~A.~Penson}
\email{penson@lptl.jussieu.fr}

\affiliation{Laboratoire de Physique Th\'{e}orique de la Mati\`{e}re Condens\'{e}e (LPTMC), Universit\'{e} Pierre et Marie Curie, CNRS UMR 7600, Tour 13 - 5i\`{e}me \'{e}t., Bo\^{i}te Courrier 121, 4 place Jussieu, F 75252 Paris Cedex 05, France}

\begin{abstract}
We study a number of possible extensions of the Ramanujan master theorem, which is formulated here by using methods of Umbral nature. We discuss the implications of the procedure for the theory of special functions, like the derivation of formulae concerning the integrals of products of families of Bessel functions and the successive derivatives of Bessel type functions. We stress also that the procedure we propose allows a unified treatment of many problems appearing in applications, which can formally be reduced to the evaluation of exponential- or Gaussian-like integrals.   
\end{abstract}

\maketitle


\section{Introduction}

The Ramanujan master theorem (RMT) \cite{GHHardy40, TAmdeberhan09, TAmdeberhan11} states that if the function $f(x)$ is defined through the series expansion 
\begin{equation}\label{eq1}
f(x) = \sum_{r = 0}^{\infty} \frac{(-x)^r}{r!} \, \varphi(r)
\end{equation} 
with $\varphi(0) \neq 0$ then the following identity holds
\begin{equation}\label{eq2}
\int_{0}^{\infty}\, x^{\nu -1}\, f(x) \, dx \,=\, \Gamma(\nu)\, \varphi(-\nu).
\end{equation}
A proof, albeit not rigorous, of the theorem can be achieved by the use of the Umbral methods \cite{SRoman84}, namely by setting \cite{TAmdeberhan11, DBabusci11}
\begin{equation}\label{eq3}
f(x) \,=\, \sum_{r = 0}^{\infty} \frac{\left(-x\right)^r}{r!}\, \hat{c}^{\,r} \varphi(0),
\end{equation}
with the operator $\hat{c}$ defined by:
\begin{equation}\label{eq4}
\hat{c}^{\,r}\, \varphi(0) \,=\, \varphi(r).
\end{equation}  
In this way the function $f(x)$ can be formally written as an pseudo-exponential function, $f(x) = e^{-\hat{c}\, x} \varphi(0)$, and thus the integral in Eq.~(\ref{eq2}) can be given in the form
\begin{equation}\label{eq5}
\int_{0}^{\infty} x^{\nu-1}\, e^{- \hat{c}\, x}\, dx\, \varphi(0) \,=\, \Gamma(\nu)\, \hat{c}^{-\nu}\,\varphi(0),
\end{equation}
and using Eq.~(\ref{eq4}) we end up with the result quoted in Eq.~(\ref{eq2}). For further comments see the concluding Section and for a rigorous proof Refs.~\cite{GHHardy40, TAmdeberhan09}.

An immediate consequence of the theorem is the evaluation of the following integral
\begin{equation}\label{eq7}
I_{\nu, \,\alpha} \,=\, \int_{0}^{\infty} x^{\nu-1}\, C_{\alpha}(x)\, dx,
\end{equation}
where
\begin{equation}\label{eq8}
C_{\alpha}(x) \,=\, \sum_{r = 0}^{\infty} \frac{(-x)^r}{r!\, \Gamma(r + \alpha + 1)}
\end{equation}
is the Tricomi-Bessel function of order $\alpha$ \cite{FGTricomi54} which satisfies the conditions of the RMT, with $\varphi(0) = 1/\Gamma(\alpha + 1)$. According to Eq.~(\ref{eq2}), we get
\begin{equation}\label{eq9}
I_{\nu, \alpha} \,=\, \frac{\Gamma(\nu)}{\Gamma(\alpha - \nu + 1)}.
\end{equation}
The procedure we have just quoted, which traces back to Crofton \cite{TAmdeberhan11} and to other operationalists, is by no means a proof of the theorem but just a guiding tool, which will be proved to be very useful for the forthcoming speculations.

As the function $C_{\alpha}(x)$ is linked to the ordinary Bessel functions by the identity
\begin{equation}\label{eq10}
C_{\alpha}(x) \,=\, (2\,x)^{-\alpha/2}\, J_{\alpha}(2\sqrt{x}),
\end{equation}
the use of the RMT may be of noticeable interest for the evaluation of various integrals appearing in applications.

As noted in Ref.~\cite{DBabusci11} the Umbral formalism may be useful to make some progress towards an extension of the RMT theorem. We observe indeed that if
\begin{equation}\label{eq11}
g_{m}(x) \,=\, e^{-\hat{c}\, x^{m}} \varphi(0) \,=\, \sum_{r = 0}^{\infty} \frac{(- x^{m})^{r}}{r!}\, \varphi(r),
\end{equation}
by applying the same procedure as in Eqs.~(\ref{eq5}), we have
\begin{equation}\label{eq12}
\int_{0}^{\infty} dx\, x^{\nu - k}\, g_{m}(x) \,=\, \frac{1}{m} \, \Gamma\left(\frac{\nu + 1 - k}{m}\right) \, \varphi\left(- \frac{\nu + 1 - k}{m}\right).
\end{equation}
Just as an application of this result, we note that integrals of the type
\begin{equation}\label{eq13}
G_{0}(b) \,=\, \int_{-\infty}^{+\infty} e^{b\, x}\, g_{2}(x)\, dx,
\end{equation}
can be written as \textit{pseudo}-Gaussian integral of the form
\begin{eqnarray}\nonumber
G_{0}(b) &=& \int_{-\infty}^{\infty} e^{\,b\, x}\, e^{- \hat{c}\, x^2}\, dx\, \varphi(0) \\ [0.7\baselineskip]\nonumber
&=& \sqrt{\frac{\pi}{\hat{c}}}\, e^{\,b^2/(4\, \hat{c})}\, \varphi(0) \,=\, \sqrt{\pi}\, \sum_{r = 0}^{\infty} \frac{b^{2\, r}}{4^r\, r!} \, \hat{c}^{-r - 1/2}\, \varphi(0) \\ [0.7\baselineskip]\label{eq14}
&=& \sqrt{\pi}\, \sum_{r = 0}^{\infty} \frac{b^{2\, r}}{4^r\, r!}\, \varphi(-r -1/2).
\end{eqnarray} 
In the case of $\varphi(r) \,=\, 1/r!\,$, Eq.~(\ref{eq14}) becomes 
\begin{eqnarray}\nonumber
G_{0}(b) &=& \int_{-\infty}^{\infty} e^{\, b\, x}\, J_{0}(2\, x)\, dx = \sqrt{\pi}\, \sum_{r = 0}^{\infty}\, \frac{b^{2\, r}}{4^r\, r!} \, \frac{1}{\Gamma(-r + 1/2)} \\ [0.7\baselineskip] \label{eq15}
&=& \frac{2}{\sqrt{4 + b^2}}.
\end{eqnarray}
Further examples will be discussed later in this paper, which is devoted to understand a variety of consequence emerging from RMT and from its Umbral revisitation.

In Section 2 we will explore its relevance in the field of special functions, while in Section 3 we will discuss some concluding remarks.


\section{The RMT and special functions}

Let us consider the integral (\ref{eq2}) in which the function $f(x)$ is given by 
\begin{equation}\label{eq16}
f(x) \,=\, e^{-a\,x - b\,x^2}.
\end{equation}
This function satisfies the RMT conditions, since it can be expanded as follows
\begin{equation}\label{eq17}
f(x) \,=\, \sum_{n = 0}^{\infty}\, \frac{(-x)^n}{n!} \, H_{n}(a, -b),
\end{equation} 
where $H_{n}(x, y)$ are two-variable Hermite polynomials defined by \cite{PAppell26, GDattoli05}
\begin{equation}\label{eq18}
H_{n}(x, y) \,=\, n!\, \sum_{r=0}^{[n/2]}\, \frac{x^{n - 2r}\, y^{r}}{(n - 2r)! \, r!}.
\end{equation}
In this case, $\varphi(n) = H_{n}(a, -b)$ so that
\begin{equation}\label{eq19}
\Gamma(\nu; a, b) \,=\, \int_{0}^{\infty} x^{\nu-1}\, e^{-a\, x - b\, x^2}\, dx \,=\, \Gamma(\nu)\, H_{-\nu} (a, -b),
\end{equation}
where
\begin{equation}\label{eq20}
H_{-\nu}(a, -b) \,=\, \Gamma(-\nu+1)\, \sum_{r=0}^{\infty} \frac{a^{-\nu - 2 r}\, (-b)^r}{\Gamma(-\nu - 2r + 1)\, r!}
\end{equation}
are no more polynomials but functions, converging for $\left\vert \, b/a^2 \right\vert < 1$.

We have denoted the integral in Eq.~(\ref{eq19}) by $\Gamma(\nu; a, b)$ to stress that it can be considered as a generalization of the Euler's gamma function. We get indeed 
\begin{equation}\label{eq21}
\Gamma(\nu; 1, 0) \,=\, \Gamma(\nu).
\end{equation}
The following properties should be also noted
\begin{equation}\label{eq22}
\partial_{a} \Gamma(\nu; a, b) \,=\, -\Gamma(\nu + 1; a, b), \qquad
\partial_{b} \Gamma(\nu; a, b) \,=\, -\Gamma(\nu + 2; a, b),
\end{equation}
which, once collected together, yield
\begin{equation}\label{eq23}
\partial_{\,b} \Gamma(\nu; a, b) \,=\, - \partial_{\,a}^{\,2} \Gamma(\nu; a, b), \qquad \text{where} \quad \Gamma(\nu; a, 0) \,=\, \frac{\Gamma(\nu)}{a^{\nu}}.
\end{equation}
From Eqs.~(\ref{eq21})-(\ref{eq23}) one obtains
\begin{eqnarray}\nonumber
\Gamma(\nu-1; a, b) &=& e^{-b\, \partial_{a}^{2}}\, \frac{\Gamma(\nu - 1)}{a^{\nu - 1}} \,=\, \frac{e^{-b\, \partial_{a}^{2}}}{\nu - 1}\, a\, \frac{\Gamma(\nu)}{a^{\nu}} \,=\, \frac{e^{-b\, \partial_{a}^{2}}\, a\, e^{b\, \partial_{a}^{2}}}{\nu-1}\, \Gamma(\nu; a, b) \\ [0.7\baselineskip] \label{eq50}
&=& \frac{1}{\nu - 1} (a - 2\, b\, \partial_{a})\, \Gamma(\nu; a, b)
\end{eqnarray}
i. e.
\begin{equation}\label{eq51}
(\nu - 1)\, \Gamma(\nu - 1; a, b) \,=\, a\, \Gamma(\nu; a, b) + 2\, b\, \Gamma(\nu + 1; a, b)
\end{equation}
which reduces to $\Gamma(\nu) = (\nu-1)\Gamma(\nu-1)$ for $a = 1$ and $b=0$.

Let us note that for the following generalization of integral (\ref{eq19})
\begin{equation}\label{eq24}
\Gamma(\nu; a, b|\, m) \,=\, \int_{0}^{\infty} x^{\nu-1}\, e^{-a\, x - b\, x^{m}}\, dx \,=\,  \Gamma(\nu)\, H_{-\nu}^{(m)}(a, -b)
\end{equation}
with 
\begin{equation}\nonumber
H_{n}^{(m)}(x, y) \,=\, n!\, \sum_{r=0}^{[n/m]}\, \frac{x^{n - m r}\, y^{r}}{(n - mr)!\, r!}
\end{equation}
higher order Hermite polynomials \cite{PAppell26, GDattoli05}, the use of RMT is not helpful. We cannot indeed apply the fact that
\begin{equation}\nonumber
e^{-a\, x - b\, x^{m}} \,=\, \sum_{n = 0}^{\infty} \frac{(-x)^n}{n!}\, H_{n}^{(m)}(a, -b),
\end{equation}
(note that the $H_{n}(x, y)$ is in fact $H_{n}^{(2)}(x, y)$), because the analogue of the series (\ref{eq20}) does not converge for $m \geq 3$. The integral in Eq.~(\ref{eq24}) is however well defined and it can be used as an independent definition of either the higher order gamma function $\Gamma(\nu; a, b|\, m)$ or of higher Hermite functions $H_{-\nu}^{(m)}(a, b)$. 

We have noted that the straightforward use of the RMT in the case of Eq.~(\ref{eq24}) leads to a diverging series. As an alternative, we consider the case
\begin{equation}\label{eq52}
\int_{0}^{\infty} x^{\nu - 1}\, e^{- \hat{c}\, b\, x^{m}}\, e^{- a\, x}\, dx \, \varphi(0) = \Gamma(\nu)\, H_{-\nu}^{(m)}(a, - b\, \hat{c})\, \varphi(0),
\end{equation}
where, according to Eq.~(\ref{eq20}),
\begin{equation}\label{eq53}
H_{-\nu}^{(m)}(a, -b\, \hat{c})\, \varphi(0) = \Gamma(1 - \nu) \sum_{r = 0}^{\infty} \frac{a^{-\nu - m\, r}\, (-b)^r}{\Gamma(-\nu - m\, r + 1)\, r!} \, \varphi(r).
\end{equation}
In the case $\varphi(r) = 1/r!\,$ the series is converging. 

A comment is also in order on the nature of the Hermite-like polynomials specified by the generating function
\begin{equation}\label{eq54}
e^{a\, x + b\, \hat{c}\, x^{m}}\, \varphi(0) \,=\, \sum_{n = 0}^{\infty} \frac{x^{n}}{n!}\, H_{n}^{(m)}(a, b\, \hat{c})\, \varphi(0).
\end{equation}
In the case of $m = 2$, $\varphi(r) \,=\, 1/r!\,$, they reduce to
\begin{equation}\label{eq55}
\tilde{H}_{n}(a, b) \,=\, H_{n}(a, b\, \hat{c})\, \varphi(0) \,=\, n! \sum_{r = 0}^{[n/2]} \frac{a^{n - 2r}\, b^{r}}{(n - 2r)!\, (r!)^2}.
\end{equation}
They are the so-called hybrid polynomials \cite{GDattoli04-1}, since they share the properties of the Laguerre and Hermite polynomials. The operational identity
\begin{equation}\label{eq56}
\tilde{H}_{n}(a, b) \,=\, e^{\,\hat{c}\, \partial_{x}^{2}}\, x^{n}\, \varphi(0) \,=\, J_{0}(i\, 2\, \partial_{x})\, x^{n}
\end{equation}
can also be used to define them.

Analogous results, concerning the Laguerre polynomials, can be now obtained by considering the integral
\begin{equation}\label{eq26}
\Lambda(\nu; a, b) \,=\, \int_{0}^{\infty} x^{\nu-1}\, e^{-a\, x}\, C_{0}(b\, x)\, dx.
\end{equation}
Since the generating function of the two-variable Laguerre polynomials writes \cite{GEAndrews01}
\begin{equation}\label{eq27}
\sum_{n=0}^{\infty} \frac{t^{n}}{n!}\, L_{n}(x, y) \,=\, e^{y\,t} C_{0}(x\, t), \qquad L_{n}(x, y) \,=\, n!\, \sum_{r=0}^{\infty} \frac{y^{n-r}\, (-x)^r}{(n-r)!\, (r!)^2},
\end{equation}
a straightforward application of the RMT yields
\begin{equation}\label{eq28}
\Lambda(\nu; a, b) \,=\, \Gamma(\nu)\, L_{-\nu}(-b, a), \qquad L_{-\nu}(-b, a) \,=\, \frac{\Gamma(1-\nu)}{a^{\nu}}\, \sum_{r = 0}^{\infty} \frac{(b/a)^{r}}{\Gamma(1-\nu - r)\, (r!)^2}.
\end{equation}

The function $\Lambda(\nu; a, b)$ too can be viewed as a generalization of the Euler's gamma function and satisfies the following recurrences
\begin{equation}\label{eq29}
\partial_{a}\, \Lambda(\nu; a, b) \,=\, - \Lambda(\nu + 1; a, b), \qquad \partial_{b} \,b\, \partial_{b}\, \Lambda(\nu; a, b) \,=\, - \Lambda(\nu + 1; a, b).
\end{equation}
Thus getting
\begin{equation}\label{eq30}
\partial_{a}\, \Lambda(\nu; a, b) \,=\, \partial_{b}\, b\, \partial_{b}\, \Lambda(\nu; a, b), \qquad \Lambda(\nu; a, 0) \,=\, \frac{\Gamma(\nu)}{a^{\nu}}.
\end{equation}
Operators of the type $\partial_{x}\, x\, \partial_{x}$ appearing in Eq.~(\ref{eq30}) are known in the literature as the Laguerre derivative \cite{GDattoli04, KAPenson09}. 

Before closing this Section we will discuss a further way of employing the RMT. We consider the integral
\begin{equation}\label{eq31}
B(\nu; a, b) \,=\, \int_{0}^{\infty} x^{\nu - 1}\, e^{-a\, x}\, J_{0}(2\, \sqrt{b}\, x)\, dx.
\end{equation}
According to the discussion in the introductory Section (see Eq.~(\ref{eq14})) we can write
\begin{equation}\label{eq32}
B(\nu; a, b) \,=\, \int_{0}^{\infty} x^{\nu-1}\, e^{-a\, x}\, e^{-\hat{c}\, b\, x^2}\, dx\, \varphi(0)
\end{equation}
and, taking into account Eq.~(\ref{eq19}) we obtain the closed-form:
\begin{equation}\label{eq33}
B(\nu; a, b) \,=\, \Gamma(\nu; a, b\,\hat{c})\, \varphi(0) \,=\, \Gamma(\nu)\, \Gamma(1 - \nu)\, \sum_{r = 0}^{\infty} \frac{a^{\nu - 2 r}\, (-b)^r}{\Gamma(-\nu - 2 r + 1)\, (r!)^2}.
\end{equation}
This is a further proof of the flexibility of the methods associated with the Umbral techniques proposed in the paper.


\section{Concluding remarks}

All the previous discussion is based on the tacitly assumed conjecture that the following identity holds
\begin{equation}\label{eq34}
\int_{0}^{\infty} x^{\nu - 1}\, e^{- \hat{c}\, x}\, \varphi(0)\, dx \,=\, \left(\int_{0}^{\infty} x^{\nu - 1}\, e^{- \hat{c}\, x}\, dx\,\right)\, \varphi(0).
\end{equation}

This is the critical point and the rigorous proof of the Ramanajuan theorem is based on the proof of such a lemma, which will be conjectured to be true. We have checked, \textit{a posteriori}, the validity of our results by benchmarking them with numerical tests. We can therefore state that, in the case of functions for which we have $\varphi(r) = 1/\Gamma(r + \beta + 1)$, the conjecture holds true.

In the Section 2 we have seen how the use of the pseudo-exponential function $e^{-\hat{c}\, x}\, \varphi(0)$ can be particularly useful to evaluate integrals involving special functions. We will now use a modified form of the RMT to evaluate integrals of the type
\begin{equation}\label{eq35}
I = \int_{-\infty}^{\infty}\, f(x)\, g(x)\, d x
\end{equation} 
with
\begin{eqnarray}\nonumber
f(x) &=& \sum_{r = 0}^{\infty} \frac{(-x)^r}{r!}\, \varphi(r), \qquad \varphi(0) \,\neq\, 0 \\[0.7\baselineskip] \label{eq36}
g(x) &=& \sum_{s = 0}^{\infty} \frac{(-x^2)^s}{s!}\, \sigma(s), \qquad \sigma(0) \,\neq\, 0. 
\end{eqnarray}
An extension of the RMT can be stated as it follows
\begin{equation}\label{eq37}
I \,=\, \int_{-\infty}^{\infty} e^{-\hat{c}\, x}\, e^{-\hat{d}\, x^2}\, dx\, \varphi(0)\, \sigma(0),
\end{equation}
where $\hat{c}^{r}\,\varphi(0) = \varphi(r)$ and $\hat{d}^{r}\,\sigma(0) = \sigma(r)$. By working out the Gaussian integral in Eq.~(\ref{eq37}), we obtain
\begin{equation}\nonumber
I \,=\, \sqrt{\frac{\pi}{\hat{d}}}\, \exp\left(\frac{\hat{c}^{2}}{4\, \hat{d}}\right)\, \varphi(0)\, \sigma(0) \,=\, \sqrt{\pi}\, \sum_{r = 0}^{\infty} \frac{1}{4^{r}\, r!}\, \varphi(2 r)\, \sigma(-r - 1/2),
\end{equation}
which holds only if the series converges. An example is provided by
\begin{eqnarray}\nonumber
\int_{-\infty}^{\infty} J_{0}(2\, \sqrt{a\, |x|})\, J_{0}(2\, b\, x)\, dx &=& \sqrt{\frac{\pi}{\hat{d}\, b^{2}}} \, \exp\left(\frac{\hat{c}^2\, a^2}{4\, \hat{d}\, b^2}\right)\, \varphi(0)\, \sigma(0) \\[0.7\baselineskip] \nonumber
&=& \frac{\sqrt{\pi}}{b} \sum_{r = 0}^{\infty}  \frac{a^{2\, r}\, b^{-2\, r}}{4^r\, r!}\, \frac{1}{\Gamma(2 r + 1)\, \Gamma(-r + 1/2)} \\[0.7\baselineskip] \label{eq38}
&=& \frac{1}{b}\, J_{0}\left(\frac{a}{2\, b}\right),
\end{eqnarray}
where $\varphi(r) = \sigma(r) = 1/\Gamma(r+1)$.

We have also used the pseudo-Gaussian function $e^{-\hat{c}\, x^2}\, \varphi(0)$ as a tool to derive integrals involving the Bessel functions. In the following we will see how the fairly wild use of the operator $\hat{c}$ may provide quite a unique tool, to study further properties of these functions. For example, we can establish formulae for the repeated derivatives of the Bessel functions by using the same methodology adopted in the case of the ordinary Gaussian function.

In the case of successive derivatives of the $0$-th order Bessel function, we set
\begin{equation}\label{eq40}
(-1)^n\, \partial_{x}^{n} J_{0}(2\, x) \,=\, (-1)^n\, \partial_{x}^{n} e^{-\hat{c}\, x^2}\, \varphi(0)
\end{equation}
and, recalling that in the ordinary case we have \cite{GDattoli97}
\begin{equation}\label{eq41}
(-1)^n\, \partial_{x}^{n} e^{-a\, x^2} \,=\, H_{n}(2 a x,\, -a)\, e^{-a\, x^2},
\end{equation}
we get
\begin{equation}\label{eq42}
(-1)^n\, \partial_{x}^{n} J_{0}(2\, x) \,=\, H_{n}(2\hat{c} x, -\hat{c})\, e^{-\hat{c} x^2}\, \varphi(0),
\end{equation}
where $\varphi(r) = 1/r!\,$. By using the explicit form of the two-variable Hermite polynomials and by expanding the pseudo-exponential, we find: \cite{YuABrychkov08}
\begin{eqnarray}\nonumber
(-1)^n\, \partial_{x}^{n} J_{0}(2\,x) &=& n!\, \sum_{r = 0}^{[n/2]}\, \frac{(-1)^r\,(2 x)^{n - 2r}}{(n-2 r)!\, r!}\, \sum_{p=0}^{\infty}\, \frac{(-1)^p\, x^{2 p}}{p!}\, \hat{c}^{\,p + n - r}\, \varphi(0) \\ [0.7\baselineskip] \label{eq43}
&=& n! \, \sum_{r = 0}^{[n/2]}\, \frac{(-1)^r\, 2^{n - 2 r}}{x^r\, (n - 2 r)!\, r!} \, J_{n-r}(2\, x)
\end{eqnarray}
and new formulae, like
\begin{eqnarray}\nonumber
\partial_{x}^{n} \left[e^{a\, x}\, J_{0}(2 \sqrt{b}\, x)\right] &=& e^{a\, x} \sum_{r = 0}^{[n/2]} \frac{(-1)^{n-r}\, n!}{r!}\, \sum_{s = 0}^{n - 2 r} \frac{(-a)^s\, b^{n - r - s}\, (2\,x)^{n - 2 r - s}}{s!\, (n - 2 r - s)!} \, \hat{c}^{n - r - s}\, e^{-b\, \hat{c}\, x^{2}}\, \varphi(0), \\ [0.7\baselineskip] \label{eq34a}
&=& e^{a\, x} \sum_{r = 0}^{[n/2]} \frac{(-1)^{n-r}\, n!}{r!\, x^{r}} \sum_{s = 0}^{n - 2 r} \frac{(-a)^s\, b^{(n - r - s)/2}\, 2^{n - 2 r - s}}{s!\, (n - 2 r - s)!}\, J_{n - r - s}\left(2\, \sqrt{b}\, x\right)
\end{eqnarray}
and
\begin{eqnarray}\nonumber
\partial_{x}^{n} \left[J_{0}(2\, \sqrt{a}\, x)\, J_{0}(2\, \sqrt{b}\, x)\right] &=& \sum_{r = 0}^{[n/2]} \frac{(-1)^{n-r}\, 2^{n - 2 r}\, n!}{(n - 2 r)!\, r!}\, x^{-r}\, \sum_{s = 0}^{n - r} \binom{n-r}{s}\, \frac{b^{s/2}}{a^{(r+s-n)/2}} \\[0.7\baselineskip] \label{eq34b}
& &\qquad\qquad\qquad\qquad \times J_{n - r- s}(2\, \sqrt{a}\, x)\, J_{s}(2\, \sqrt{b}\, x)
\end{eqnarray}
which are quite interesting results, difficult to get by conventional means and absent in available collections \cite{YuABrychkov08}. This is just an example and other will be discussed in a forthcoming investigation. 

Most of the considerations developed in this paper have been based on the properties of the pseudo-exponential function $e^{-\hat{c}\, x}\, \varphi(0)$ which, as for the ordinary case, possesses the semi-group property
\begin{equation}\label{eq44}
e^{\hat{c}\, (x + y)} \,=\, e^{\hat{c}\, x}\, e^{\hat{c}\, y}
\end{equation}
only if $x$ and $y$ are commuting quantities. If this is not true and if, for example, $\hat{X}$ and $\hat{Y}$ are operators such that
\begin{equation}\label{eq45}
\left[\hat{X}, \, \hat{Y}\right] \,=\, \hat{\mathbb{I}}\, k,
\end{equation}
where $k$ is a $c$-number and $\hat{\mathbb{I}}$ the unit operator, we can obtain a convenient disentanglement by using the Weyl-identity \cite{GDattoli05}, namely
\begin{equation}\label{eq48}
e^{\hat{c}\, \left(\hat{X} + \hat{Y}\right)} \,=\, e^{\hat{c}\, \hat{X}}\, e^{\hat{c}\, \hat{Y}}\, e^{-k\, \hat{c}^2/2}.
\end{equation}
If $\hat{X} \,=\, k\,x$, $\hat{Y} \,=\, \partial_{x}$ we obtain
\begin{equation}\label{eq48}
e^{\hat{c}\,(\hat{X} + \hat{Y})}\, \varphi(0) = \sum_{r=0}^{\infty}\, \frac{(k\,x)^r}{r!}\, \sum_{s=0}^{\infty}\, \frac{(-k)^s}{2^s\, s!}\, \hat{c}^{\, 2 s + r}\, \varphi(0) \,=\, \sum_{r = 0}^{\infty} \frac{(k\, x)^{r}}{r!}\, W_{r}(-k/2 |\, 2),
\end{equation}
where 
\begin{equation}\label{eq48a}
W_{r}(x |\, \nu) = \sum_{n = 0}^{\infty} \frac{x^n}{n!\, \Gamma(\nu\, n + r + 1)}
\end{equation}
is the Bessel-Wright function \cite{GEAndrews01}.

The above remark is a further example of how rich may be the consequences offered by the method we are proposing. In a forthcoming paper we will see their importance in the study of partial differential equations of evolution type, involving non standard derivative operators.

Finally we want to mention the relevance of the presented approach for the theory of Hankel's integral transform \cite{LDebnath07, APPrudnikov92}, which can be interpreted as
\begin{equation}\label{eq57} 
\left(H_{\alpha}\, f\right)(x) \,=\, \int_{0}^{\infty} f(t)\, J_{\alpha}(2\, x\, t)\, dt = \int_{0}^{\infty} f(t)\, e^{\,- \hat{c}\, x^{2}\, t^{2}}\, dt\, \varphi(0)
\end{equation}
i. e., according to the point of view developed in this paper, as pseudo-Gaussian integrals. Further comments on this last topic will be presented elsewhere.


\ \\

The authors acknowledge support from Agence Nationale de la Recherche (Paris, France) under Program PHYSCOMB No. ANR-08-BLAN-0243-2. G.~Dattoli thanks the University Paris XIII for financial support and kind hospitality.


\end{document}